\pdfoutput=1
\documentclass[12pt]{article}
\usepackage[utf8]{inputenc} 
\usepackage[T1]{fontenc}
\usepackage{hyperref}
\usepackage{url}      
\usepackage{setspace}
\usepackage[pdftex]{graphicx}
\usepackage{dcolumn}
\usepackage[margin=1in]{geometry}
\usepackage{amsmath}
\usepackage{graphicx}
\usepackage{float}
\usepackage{tabularx}
\usepackage[sort, numbers]{natbib}
\usepackage[mathscr]{eucal}
\usepackage{multirow}
\usepackage{booktabs}
\usepackage{xcolor}

\usepackage{caption}
\usepackage{setspace}

\providecommand{\keywords}[1]
{
  \small	
  \textbf{\textit{Keywords---}} #1
}

\begin{document}

\title{Constraining logarithmic $f(R,T)$ model using Dark Energy density parameter $\Omega_{\Lambda}$ and Hubble parameter $H_0$}

\author{ Biswajit Deb  \footnote{Electronic Address:biswajitdeb55@gmail.com} , Atri Deshamukhya  \footnote{Electronic Address: atri.deshamukhya@aus.ac.in}\\
Department of Physics, Assam University, Silchar, India }

\date{}
\maketitle

\begin{abstract}
Of many extended theories of gravity, $f(R,T)$ gravity has gained reasonable interest in recent times as it provides interesting results in cosmology. Logarithmic corrections in modified theories of gravity has been studied extensively. In this work, we considered logarithmic correction to the trace term T and take the functional form as $f(R,T)=R + 16 \pi G \alpha \ln T$ where $\alpha$ is a free parameter. The free parameter is constrained using dark energy density parameter $\Omega_{\Lambda}$ and Hubble parameter $H_0$. The lower bound is found to be $\alpha \ge - 9.85 \times 10^{-29}$. The cosmological implications are also studied. 

\end{abstract} \hspace{8pt}


\section{Introduction}	
The remarkable discovery in 1998 by Riess et. al and Perlmutter et. al paused a fundamental question on the classical General Relativity (GR) \cite{1,2}. The standard GR predicts that the expansion of the Universe should be decelerating with time. Instead, it is observed that the current Universe has entered a second phase of accelerated expansion which started around $z=1$ \cite{1,2}. Besides this, it is observed that nearly 70\% of the total energy density of the Universe is in some mysterious form called Dark Energy \cite{1,2}. Whether this Dark Energy is driving the present expansion of the Universe is a matter of investigation.\\
\\
Einstein field equations,  in their original form can't explain this late time acceleration because it gives an equation of state (EoS) parameter $w=0$ \cite{3}. Whereas, the present late time acceleration demands EoS parameter to be $w< - \frac{1}{3}$. Now, this can be explained if one considers the Universe filled with some exotic form of fluid with EoS parameter $w< - \frac{1}{3}$ \cite{3}. That means the dominant Dark Energy constituent is driving the expansion. Apart from this, cosmological constant with EoS parameter $w=-1$ is a viable alternative explanation for Dark Energy as well as late time expansion of the Universe \cite{3}. In fact the $\Lambda$CDM model gives the best fit results to the present observed Universe \cite{4}. But the Cosmological constant $\Lambda$ is into fine tuning problem \cite{5}. \\
\\
$f(R)$ theory, a class of modified gravity, emerged in 1980's and gained popularity as it can explain the early inflationary Universe without considering the scalar fields in the theory \cite{6}. Since then different modifications of gravity viz. $f(G)$, $f(R,G)$, $f(G,T)$, $f(R,T)$, $f(Q)$, $f(Q,T)$ etc. have been proposed and studied in literature where R, Q, G and T are the Ricci scalar, non-metricity, Gauss-Bonnet scalar and Trace of the Energy momentum tensor respectively \cite{7,8,9,10,11,12}. After the discovery of late time acceleration, studies of modified theories of gravity gained even more interest as it has the potential to explain late time acceleration of the Universe without invoking Dark Energy in the theory \cite{13}. \\
\\
$f(R,T)$ theory of gravity, proposed by Harko et. al is found to be extremely sound in explaining cosmological phenomena. It has been studied with reference to inflation \cite{14,15,16,17}, dark energy \cite{18,19,20,21,22}, dark matter \cite{23}, wormhole \cite{24,25,26,27,28,29,30}, pulsar \cite{31,32}, white dwarfs \cite{33}, gravitational waves \cite{34,35}, scalar field models \cite{36,37}, anisotropic models \cite{38,39}, bouncing cosmology \cite{40,41}, big-bang neucleosysnthesis \cite{42}, baryogenesis \cite{43}, brane world \cite{44,45} etc. Further, the energy conditions and junction conditions in $f(R,T)$ gravity have also been studied \cite{46,47,48,49}. Snehasish et. al developed a novel way to impose lower bound on the model parameter $\lambda $ of the simplest $f(R,T)=R+2\lambda T$ model through the equation relating the cosmological constant and critical density of the universe \cite{18}. This method can be applied to other complex forms of $f(R,T)$ to constrain the model parameter(s). This might reveal interesting result. \\
\\
Logarithmic correction in modified gravity theories has been studied extensively. The first Logarithmic correction to trace term T in $f(R,T)$ theory has been proposed by Elizalde et. al where they have studied the stability conditions \& energy conditions of the model \cite{50}. In this work, we have considered the simplest $f(R,T)$ model with logarithmic correction to trace term T as $f(R,T)= R+ 16\pi G \alpha \ln T$, where $\alpha$ is the model parameter and will constrain the model parameter $\alpha$ using equation relating to the cosmological constant $\Lambda$ and critical density of the universe $\rho_{cr}$. Recently Maurya et. al have studied similar model and found that it shows a quintessence dark energy model and late time universe approaches to $\Lambda$CDM model \cite{51}.\\
\\
The paper has been organised as follows: In section 2, we present a brief review of $f(R,T)$ gravity. In section 3, we present mathematical framework to impose bound on the model parameter. In section 4, we present our conclusion. Here, we will use (+,-,-,-) sign convention for the metric tensor.

\section{A brief note on $f(R,T)$ gravity}

In $f(R,T)$ gravity, the gravitational Lagrangian depends on a general function of Ricci scalar $R$ as well as of the trace of energy momentum tensor $T$. The action in ${f(R,T)}$ gravity reads as \cite{10},

\begin{equation}
	S = \int \left[\frac{f(R,T)}{16 \pi G} + L_m  \right] \sqrt{-g} d^4x 
\end{equation}
where $L_m$ is the matter Lagrangian, g is the metric determinant and G is the Newtonian gravitational constant. On variation of the action with respect to the metric, we obtain the modified field equations as,
	\begin{equation}
		f_R (R,T) R_{\mu\nu} - \frac{1}{2} g_{\mu\nu}f(R,T) + [g_{\mu\nu} \nabla_\sigma \nabla^\sigma - \nabla_\mu \nabla_\nu] f_R (R,T) =  8\pi G T_{\mu\nu} - f_T (R,T) (T_{\mu\nu} + \Theta_{\mu\nu})
	\end{equation} 
where we have denoted $f_R (R,T)= \frac{\partial f(R,T)}{\partial R}$ , $f_T (R,T)= \frac{\partial f(R,T)}{\partial T}$ and defined $T_{\mu\nu}$ and $\Theta_{\mu\nu}$ as,
\begin{equation}
	T_{\mu\nu} = g_{\mu\nu}L_m - 2 \frac{\delta L_m}{\delta g^{\mu\nu}}
\end{equation}
\begin{equation}
	\Theta_{\mu\nu}= g^{\beta\gamma} \frac{\delta T_{\beta\gamma}}{\delta g^{\mu\nu}} = -2 T_{\mu\nu} + g_{\mu\nu}L_m - 2 g^{\beta\gamma} \frac{\delta^2 L_m}{\delta g^{\mu\nu} \delta g^{\beta\gamma}}
\end{equation}
The term $\Theta_{\mu\nu}$ plays a significant role in ${f(R,T)}$ gravity. Since it contains matter Lagrangian $L_m$, depending on the nature of the matter field, the field equation for ${f(R,T)}$ gravity will vary. Besides this the functional form of ${f(R,T)}$ will also change the field equation. Thus, the field equations in ${f(R,T)}$ gravity depend both on the nature of matter field and choice of the functional form of $f(R,T)$. \\ \\
Harko proposed four different forms of the function ${f(R,T)}$ in his paper\cite{10}, viz.:
\begin{itemize}
	\item $f(R,T) = R + 2 f(T)$
	\item $f(R,T) = f_1(R) + f_2(T)$
	\item $f(R,T) = f_1(R) + f_2(R)f_3(T)$
	\item $f(R,T^{\phi}) = R + f(T^{\phi})$, where $\phi$ is a self interacting scalar filed.
\end{itemize}
Beside these forms, various other forms have been proposed in literature \cite{15, 52, 53} which are as follows:
\begin{itemize}
	\item $f(R,T) = R + \phi f(T)$, $\phi$ is scalar field.
	\item $f(R,T) = R + \alpha e^{RT}$, $\alpha$ is a constant.
	\item $f(R,T) = R + \alpha e^{\beta T} + \gamma T^n$, $\alpha, \beta, \gamma$ are constant.
\end{itemize}


\section{Constraining model parameter}
Considering $f(R,T)= R+ 16\pi G \alpha \ln T$, the Einstein-Hilbert action becomes,
\begin{equation}
	S = \int \sqrt{-g} \left[\frac{R}{16 \pi G} +  \alpha \ln T + L_m \right]  d^4x 
\end{equation}
This leads to field equations,
\begin{equation}
	R_{\mu\nu} - \frac{1}{2}g_{\mu\nu}R = 8 \pi G T_{\mu\nu}^{(eff)}
	\label{2}
\end{equation}
where $T_{\mu\nu}^{(eff)}$ is the effective stress-energy tensor given by,
\begin{equation}
	T_{\mu\nu}^{(eff)}= T_{\mu\nu} - \frac{2\alpha}{T}\left( T_{\mu\nu} - \frac{T}{2} g _{\mu\nu}\ln T + \Theta_{\mu\nu}\right)
\end{equation}
Clearly, depending on the nature of the matter field, the field equation for $f(R,T)$ gravity will be different. Now, assuming the Universe is dominated by perfect fluid, the energy-momentum tensor is
\begin{equation}
	T_{\mu\nu}=(\rho+p)u_{\mu}u_{\nu}-pg_{\mu\nu}
\end{equation}
and hence the matter Lagrangian density can be assumed as $L_m=-p$. Now, let us assume the Friedmann-Lemaitre-Robertson-Walkar (FLRW) metric in spherical coordinate for flat Universe,
\begin{equation}
	ds^2= dt^2 - a(t)^2 \left[dr^2+r^2(d\theta^2 + \sin^2\theta d\phi^2)\right]
\end{equation}
where a(t) denotes scale factor of the Universe. Now, considering the FLRW metric, the 00 component of field Eq. \ref{2} yields first modified Friedmann equation as,
\begin{equation}
	3H^2 = 8\pi G \left[\rho+\frac{2\alpha}{\rho-3p}\left(\rho+p+\frac{\rho-3p}{2}\ln (\rho-3p)\right)\right]
\end{equation}
where H is Hubble parameter, $\rho$ is energy density and p is pressure of the Universe. Further, $T=\rho-3p$ is the trace of the EM tensor. Now using equation of state parameter $w=p/\rho$, the above Friedmann equation can be rearranged as

\begin{equation}
		3H^2 = 8\pi G \left[\rho+\frac{2\alpha}{(1-3w)\rho}\left((1+w)\rho+\frac{(1-3w)\rho}{2}\ln (1-3w)\rho\right)\right]
  \label{aa}
	\end{equation}
Current observation \cite{4} suggest that $w=-1$ and hence substituting this value in Friedmann Eq. \ref{aa} yields,
\begin{equation}
	3H^2 = 8\pi G (\rho+\alpha\ln4\rho)
	\label{9}
\end{equation}
Now, Friedmann equation in GR with Cosmological Constant $\Lambda$ is given by \cite{54},
\begin{equation}
	3H^2 = 8\pi G \rho+ \Lambda c^2
	\label{10}
\end{equation}
Equating Eq.\ref{9} and Eq.\ref{10} yields,
\begin{equation}
	\Lambda c^2 = 8\pi G \alpha \ln 4 \rho
	\label{11}
\end{equation}
Now, Cosmological Constant $\Lambda$ can be defined in terms of present value of Hubble parameter $H_0$ and dark energy density parameter $\Omega_\Lambda$ as \cite{55},
\begin{equation}
	\Lambda= 3 \left(\frac{H_0}{c}\right)^2 \Omega_\Lambda
	\label{12}
\end{equation}
Substituting Eq.\ref{12} in Eq.\ref{11} we get,
\begin{equation}
	\Omega_\Lambda=\frac{8\pi G \alpha}{3H_0^2} \ln 4\rho
\end{equation}
The present Universe is spatially flat. As a result the total density parameter is $\Omega_0=\rho/\rho_{cr}=1$ \cite{4}. Hence we can replace $\rho$ by $\rho_{cr}$ in the previous equation.
\begin{equation}
	\Omega_\Lambda=\frac{8\pi G \alpha}{3H_0^2} \ln 4\rho_{cr}
	\label{14}
\end{equation}
Now the critical density of the Universe is defined as \cite{56}
\begin{equation}
	\rho_{cr}= \frac{3H_0^2}{8\pi G}
	\label{15}
\end{equation}
From Eq.\ref{14} and Eq.\ref{15} we get the final expression of the model parameter $\alpha$ in terms of Hubble parameter and dark energy density parameter as
\begin{equation}
	\Omega_\Lambda=\frac{8\pi G \alpha}{3H_0^2} \ln \frac{3H_0^2}{2\pi G}
	\label{16}
\end{equation}
From Planck 2018 data \cite{4}, we have $H_0 = 67.4 \pm 0.5 Kms^{-1}MPc^{-1}$ and $\Omega_\Lambda=0.6889 \pm 0.0056$. Substituting $H_{0}=2.17 \times 10^{-18}$ $s^{-1}$(in SI unit) and $\Omega_\Lambda=0.6833$ in Eq.\ref{16}, we obtain the lower bound on the model parameter
\begin{equation*}
	\alpha \ge - 9.85 \times 10^{-29}
\end{equation*}


\section{Conclusion}
Free parameters in modified gravity theories are trivial and hold significant role. It allows a particular gravity model to be consistent with observational results. In this work we tried to constrain the simplest $f(R,T)$ model with logarithmic correction, using Hubble parameter and dark energy density parameter. The analysis reveals that the model parameter $\alpha$ can assume any non negative value.\\
\\
This method of imposing lower bound on the model parameter by relating equation to the cosmological constant and critical density of the Universe developed by Snehasish et. al is exclusively model dependent. For more complex forms of $f(R,T)$, more input parameters are required to constrain the model. Further, this method can be applied to constrain other modified gravity models but it may require more constraining parameters beside dark energy density parameter.\\
\\
Snehasish et. al obtained the lower bound for $f(R,T)=R+2\lambda T$ model which is of the order of $10^{-8}$. In case of $f(R,T)=R+16 \pi G \alpha \ln T$, the lower bound obtained is of the order of $10^{-29}$. In both the cases, the lower bound on the model parameter is quite small. As a result, these bounds need to be validated from other sources like spectral indices, tensor-to-scalar ratio etc which is beyond the scope of this work.\\
\\
As a possible extension of this work, one can apply this methodology to other $f(R,T)$ models, specially non-minimally coupled ones. Further, this method can also be applied to other modified gravity theories like $f(R,G), f(G,T)$ etc. which may generate interesting results.

\end{document}